\documentclass[11pt,twoside]{article}

\usepackage{asp2006}
\usepackage{fancyhdr,graphicx,caption,rotating,multirow,lscape}
\DeclareGraphicsExtensions{.eps,.eps.gz,.ps,.jpg,.tiff}
\begin{document}

%%% Fill in title
\title{Radial Velocity Follow-up of Planetary Transit Candidate MACHO.120.22303.5389}  

%\title{}

%%% Fill in author names, use initials and surname and affiliation
%%% One author
%\author{J. W. Doe}
%\affil{Max-Planck Institut f\"ur Astronomie, 69117 Heidelberg, Germany}

%%% Two authors and more
\author{David T. F. Weldrake, Johny Setiawan, Patrick Weise $\&$ Thomas Henning}
\affil{Max-Planck-Institut f\"ur Astronomie, K\"onigstuhl 17, Heidelberg, 69117 Germany}

\begin{abstract}We present preliminary results on the radial velocity follow-up of a planetary transit candidate (P=2.43d, V=15.4 mag) detected during the MACHO project. The photometry is consistent with a grazing transit of an object with radius $\ge$1.8R$_{\rm{J}}$ orbiting a K dwarf star, and is the brightest best candidate detected from MACHO. Results from the 2.2m MPG/ESO telescope and FEROS (R=48,000) in May 2006 display an apparent radial velocity variation with amplitude $\sim$650m/s with the same period as the transit, and a solar-type primary. This is consistent with an orbiting companion of mass $\sim$4M$_{\rm{J}}$. However, further observations display an additional secondary long-period variation with amplitude of several km/s, indicating the presence of a third body. The system is likely a low mass eclipsing binary orbiting the solar-type primary. Further observations are planned to fully characterize the system.
\end{abstract}

\section{Project Overview}
The confirmation of transiting extrasolar planets requires a vigorous follow up process with accurate radial velocity (RV) observations. We are currently performing a feasibility study to determine the sensitivity limits of the 2.2m ESO/MPG telescope coupled with FEROS in measuring accurate ($\le$50 m/s) radial velocities of fainter transit candidates (14.0 mag$<$V$<$16.0 mag) identified from ongoing photometric surveys. 

As part of the MACHO project to identify microlensing events, the Galactic Bulge was observed over a long temporal baseline. This data was also used to search for transiting systems. \citet{DC2004} published nine candidates which best match the expected photometric signature of a transiting planet. The brightest of these candidates is MACHO.120.22303.5389 (MACHO.5389, V$=$15.3 mag, V$-$R$=$0.53 mag), with published photometry consistent with an object with radius 1.8R$_{\rm{J}}$ transiting a K-dwarf star with radius 0.86R$_{\odot}$ with a 2.43d period. Such a system would display easily visible RV variations. We have performed RV measurements of MACHO.5389 with FEROS. We present the current results of our RV analysis and describes the likely nature of this interesting system. MACHO.5389 is not simple to interpret, displaying apparent RV variations which at first glance are consistent with an orbiting planetary mass object, yet further spectroscopy also indicates the presence of a third body in the system. Likely of triple configuration, future observations with a larger telescope are planned to determine the masses of the individual components.

\section{Observations and Current Results}
\begin{figure}[!t]
\centering
\includegraphics[angle=0,width=7.2cm]{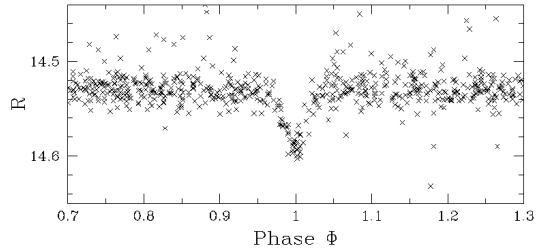}
\caption{Phase-wrapped R-band photometry for MACHO.5389 (P$=$2.43d, depth$=$0.06mag, duration$=$3.3hrs). The transit is V-shaped, indicating a grazing configuration and shows no evidence of gravity effects.\label{phot}}
\end{figure}

Fig.\space{\ref{phot}} presents the R-band photometry of MACHO.5389 as downloaded from the online MACHO photometric database\footnote{http://wwwmacho.mcmaster.ca/Data/MachoData.html}. Data in both V and R are available. The transit is V-shaped, indicating a grazing passage of the transiting object. This is perhaps indicative of a grazing binary star, yet it is expected statistically that $\sim20\%$ of transiting planets will display such a configuration. The transit does not show a secondary eclipse or out of eclipse ellipsoidal variations, and shows no discernable difference in parameters in the two filters. This indicates that the transiting companion has very little gravitational effect on the primary, having either the same color as the primary, or has negligible luminosity. 

In May 2006, we observed MACHO.5389 with the 2.2m MPG/ESO telescope and FEROS (R$=$48,000 with radial velocity precision of 10m/s for bright objects). The best spectrum (S/N$=$200) was used to determine the stellar parameters by measuring the equivalent widths of 40 FeI and 6 FeII lines, following \citet{GS1999,T2002} and \citet{J2006}. We computed the effective temperature, surface gravity and metallicity from the measured equivalent widths. We found that the primary is a G1$-$G1.5V star with M$_{\ast}=$1.0M$_{\odot}$, T$_{\rm{eff}}=$5824$\pm$70K, $\rm{[Fe/H]}=$$-$0.04$\pm$0.05 and log(g)$=$4.65$\pm$0.15.

The May observations seemingly display a radial velocity variation with amplitude $\sim$650m/s and periodicity close to the transit period (seen in Fig.\space\ref{radvels}, left). At first glance it seems that the primary has a planetary companion, responsible for the transits, with mass 4.2M$_{\rm{J}}$ and radius $\ge$2.2R$_{\rm{J}}$. This is comparable to the RV detected close companion of Tau Boo, which has a minimum mass of $\sim$4.0M$_{\rm{J}}$ \citep{B1997}. However, small-scale line asymmetries were observed, indicating the presence of a second superimposed spectrum and a possibly blended configuration for this system. The transit duration is also anomalously long (by a factor of 1.7$-$2) to be consistent with a grazing object orbiting a solar radius primary in a 2.43d orbit. The asymmetries did not change phase with the RV signal, unlike that expected for a blend, necessitating further investigation. A second set of spectroscopy was obtained in July 2006. With the two month baseline, a second large-amplitude linear trend was observed in the RV for this system, with amplitude of several km/s, as well as perhaps recovering the 650m/s variation (both seen in Fig.\space{\ref{radvels}, right). This indicates the presence of a further component, and a more complicated system configuration. 

\begin{figure}[!t]
\centering
\includegraphics[angle=0,width=6.4cm]{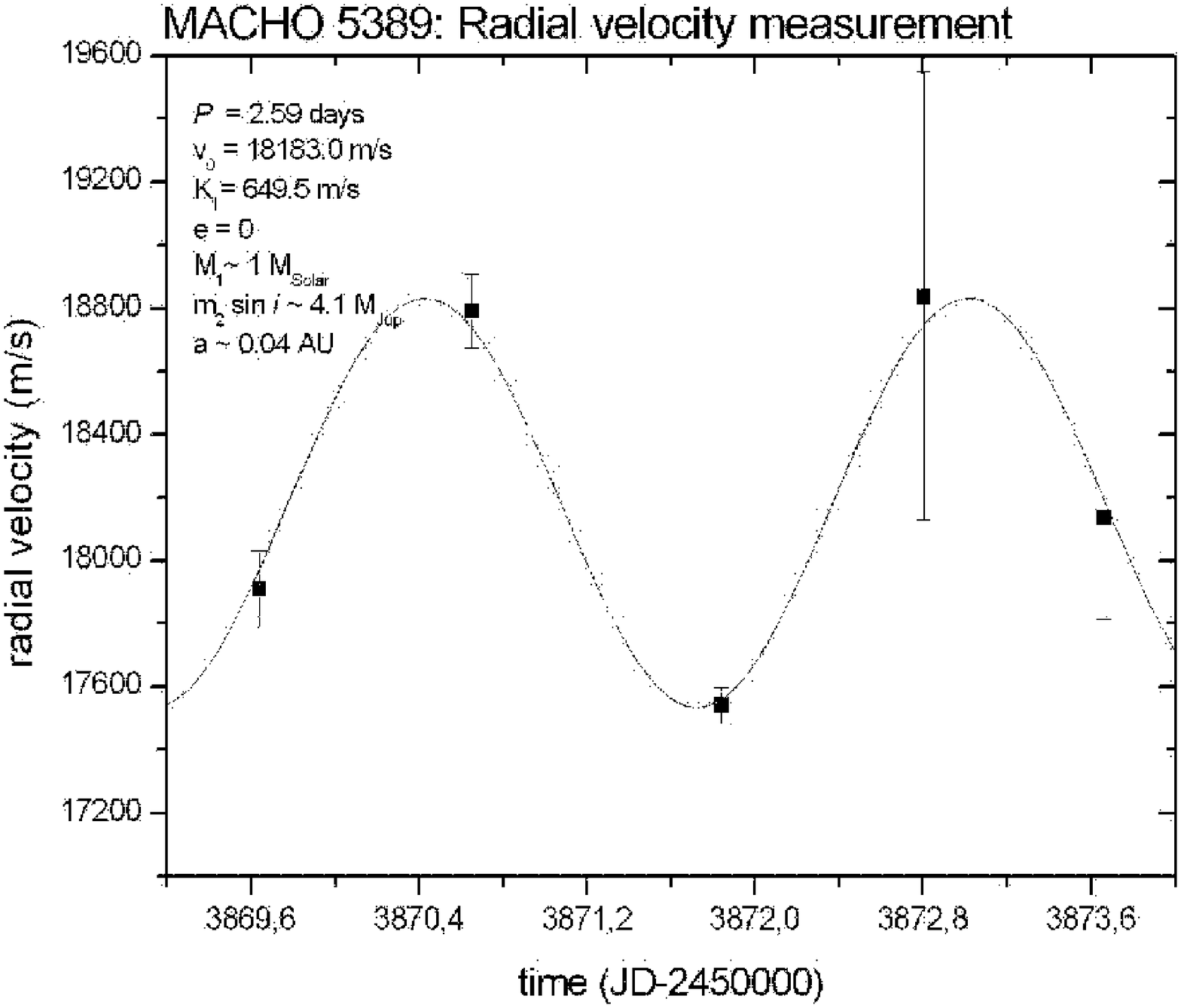}
\hspace{0.01cm}
\includegraphics[angle=0,width=6.4cm]{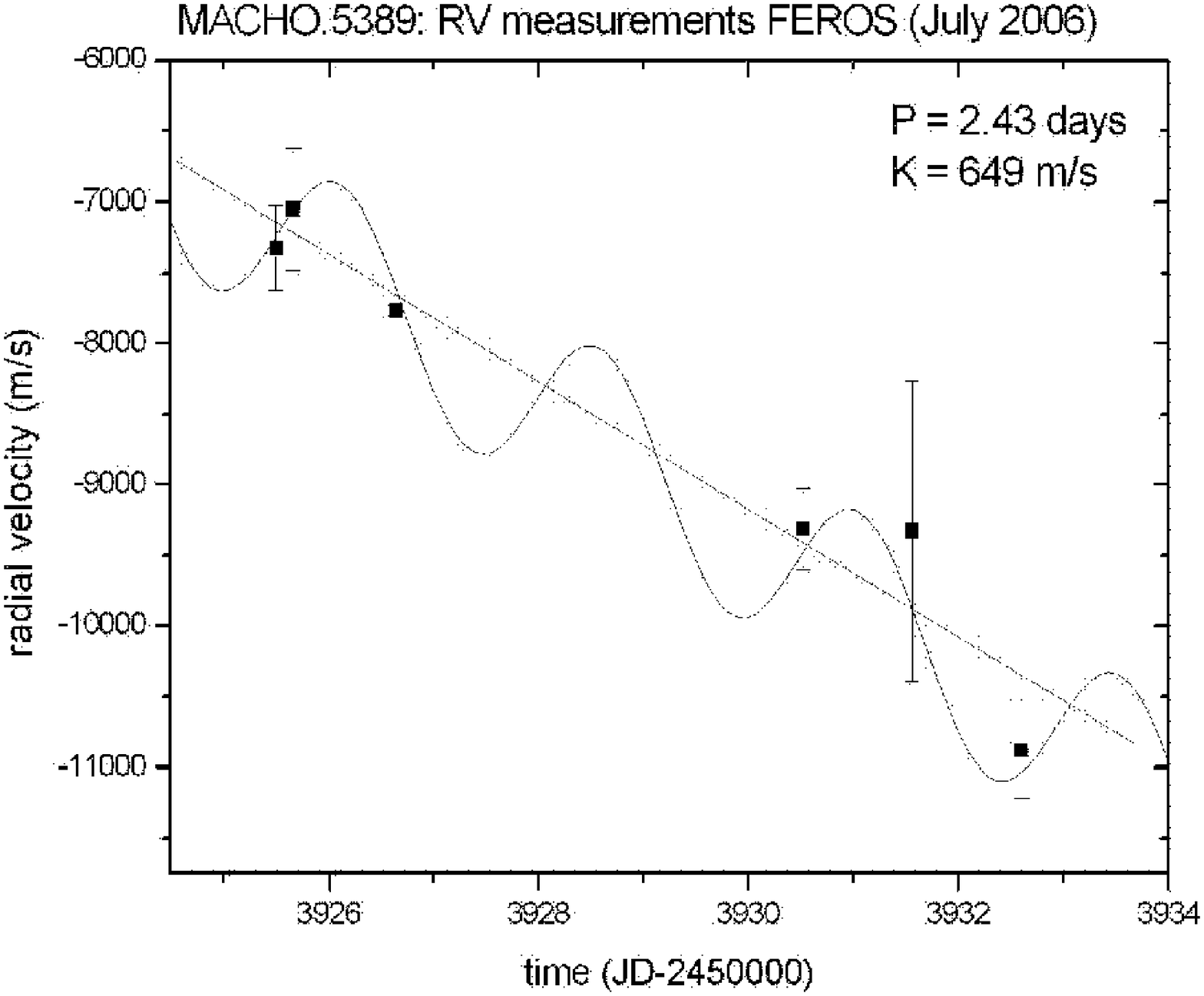}
\caption{Radial velocities for MACHO.5389, observed with the 2.2m ESO/MPG and FEROS in May (left) and July (right) 2006. The May obervations reveal an apparent variation with periodicity close to the transit period, perhaps indicative of a planetary-mass companion. The July observations display an additional large amplitude trend, revealing the presence of a third body in the system.\label{radvels}}
\end{figure}

\section{The Nature of MACHO.5389}
We envisage two scenarios. The first is more likely; that MACHO.5389 is a triple star, with a low-mass eclipsing binary with a period equal to the transit, orbiting a solar-type primary with a currently unknown period. Here,the observed 650m/s RV variation is attributed to the blended eclipsing binary, which could be in the brown dwarf mass regime. The second scenario, although somewhat unlikely from photometric analysis, is for a massive Hot Jupiter planet orbiting a solar-type primary, which is itself orbited by a close stellar companion. In this case the 650m/s variation is due to the `planet', and the long-term RV due to the stellar companion. Either scenario is of great importance to studies of the formation and survival of substellar objects. We shall use additional spectra to determine the full amplitude and period of the long-term variation, and confirm or refute the 650m/s RV amplitude seen with the similar period as the transit. By combining this with existing FEROS spectra, we can determine the total mass of the three components and the nature of this interesting system.

%\acknowledgements

%%% THE BIBLIOGRAPHY

\end{document}